\documentstyle[12pt,clustering,epsf]{article}
\def\be{\begin{equation}}

\def\ee{\end{equation}}

\def\dg{\mbox{$^\circ$}}		

\def\hMpc{h^{-1}{\rm Mpc}}
\def\h3Mpc{h^{-3}{\rm Mpc}^3}

\def\h3Mpcinv{h^{3}{\rm Mpc}^{-3}}

\def\spose#1{\hbox to 0pt{#1\hss}}
\def\simlt{\mathrel{\spose{\lower 3pt\hbox{$\mathchar"218$}}
     \raise 2.0pt\hbox{$\mathchar"13C$}}}
\def\simgt{\mathrel{\spose{\lower 3pt\hbox{$\mathchar"218$}}
     \raise 2.0pt\hbox{$\mathchar"13E$}}}

\begin{document}
\heading{THE SLOAN DIGITAL SKY SURVEY: STATUS AND PROSPECTS}

\author{Jon Loveday}
       {Fermilab, Batavia, USA.}
	{(On behalf of the SDSS collaboration.)}
\begin{center}
\leavevmode
\epsfxsize=37mm
\epsfbox{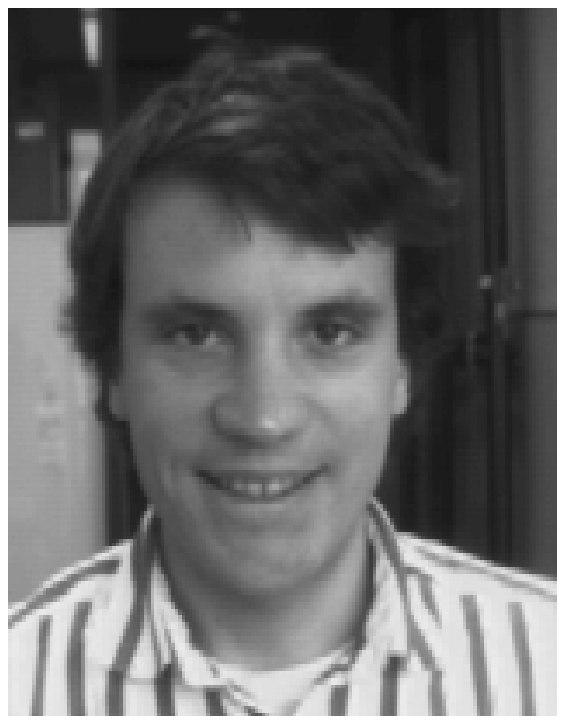}
\epsfxsize=0.6\textwidth
\end{center}

\begin{abstract}{\baselineskip 0.4cm 
The Sloan Digital Sky Survey (SDSS) is a project to definitively map $\pi$
steradians of the local Universe.  An array
of CCD detectors used in drift-scan mode will digitally image the sky
in five passbands to a limiting magnitude of $r' \sim 23$.
Selected from the imaging survey, $10^6$ galaxies
and $10^5$ quasars will be observed spectroscopically.
I describe the current status of the survey, which is due to begin
observations early in 1997, and its prospects
for constraining models for dark matter in the Universe.}
\end{abstract}

\section{Introduction}

Systematic surveys of the local Universe ($z \simlt 0.2$) can provide
some of the most important constraints on dark matter,
particularly through the measurement of the clustering of galaxies
and clusters of galaxies on large scales.
Most existing galaxy and cluster catalogues are based on photographic
plates \cite{msel90, chm89}, and there is growing concern that such surveys
might suffer from severe surface-brightness selection effects,
so that they are missing a substantial fraction of the galaxy population.
In addition, the limited volume of existing redshift surveys means
that even low-order clustering statistics, such as the galaxy two-point
correlation function, cannot reliably be measured on scales beyond
$100 \hMpc$, an order of magnitude below the scale on which COBE has
measured fluctuations in the microwave background radiation.

A collaboration has therefore been formed with the aim
of constructing a definitive map of the local universe, incorporating
digital CCD imaging over a large area in several passbands and
redshifts for around one million galaxies.
In order to complete such an ambitious project over a reasonable timescale,
it was decided to build a {\em dedicated} 2.5-metre telescope
equipped with a large CCD array imaging camera and multi-fibre spectrographs.
The collaboration comprises around 100 astronomers and engineers from
University of Chicago, Fermilab, Princeton University,
Institute for Advanced Study, Johns Hopkins University, US Naval Observatory,
University of Washington and the JPG---a group of astronomers in Japan.
The total cost of the survey is around \$30 million, and funding sources
include the Alfred P. Sloan Foundation, the National Science Foundation
and the participating institutions.

\section{Survey Overview}

The survey site is Apache Point Observatory, New Mexico, at 2800 metres 
elevation.
While better sites probably exist in Chile and atop Mauna Kea,
for a survey with such state-of-the-art instrumentation and significant on-site
manpower requirements (eg. fibre plugging and changing spectroscopic plates),
it was decided to use a site within mainland USA and with good communications
and existing infrastructure.

\begin{figure}[htbp]
\parbox{10cm}{
\epsfxsize=10cm
\epsfbox{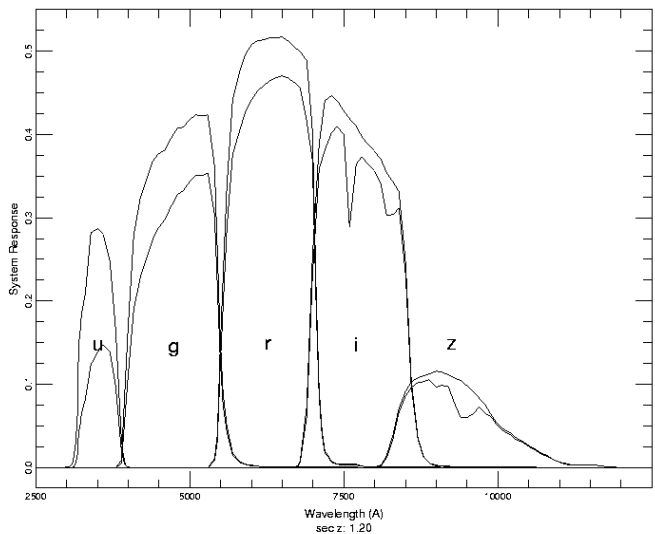}
}
\parbox{6cm}{
\caption[]{SDSS system response curves, with (lower) and without (upper)
	atmospheric extinction.}
\label{fig:filters}
}
\end{figure}

\begin{figure}[htbp]
\parbox{10cm}{
\vspace{-2cm}
\epsfxsize=10cm
\epsfbox{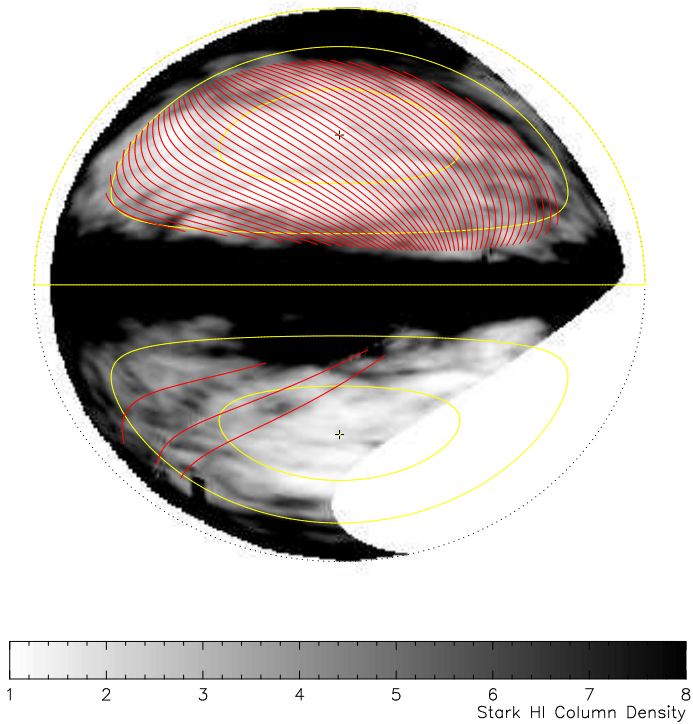}
}
\parbox{6cm}{
\caption[]{Whole sky plot showing the location of SDSS scans.
	The light lines show galactic latitudes of $b = 0$, $\pm 30$ and 
	$\pm 60\dg$, with the north and south Galactic poles being the
	upper and lower crosses respectively.
	The Galactic plane runs horizontally through the middle of the plot
	and the grey scale map shows Stark HI contours in units of
	$10^{20}$ cm$^{-2}$.
	The dark lines show the survey scan-lines, all of which follow 
	great circles.
	We observe a contiguous area of $\pi$ sr in the north, and
	three separated stripes in the south.
	Note that the northern survey is tilted with respect to the
	$b = +30\dg$ contour to avoid regions of high HI column density.}
\label{fig:stripes}
}
\end{figure}

The survey hardware comprises the main 2.5-metre telescope, equipped with
CCD imaging camera and multi-fibre spectrographs, a 0.6-metre monitor
telescope and a $10\mu$ all-sky camera.
On the best nights (new moon, photometric, sub-arcsecond seeing)
the 2.5-metre telescope will operate in imaging mode, drift scanning the
sky at sidereal rate, and obtaining nearly simultaneously images
in the five survey bands $u'$, $g'$, $r'$, $i'$ and $z'$.
The system response curves through the five filters are shown in
Figure~\ref{fig:filters}.
On sub-optimal nights, which will comprise the bulk of observing time,
the imaging camera will be replaced with a spectroscopic fibre plug-plate.
It is planned that imaging data will be reduced and calibrated, spectroscopic
targets selected, and plates drilled within the one-month lunar cycle,
so that we will be obtaining spectra of objects that were imaged the
previous month.
We will spend most of the time observing within a contiguous $\pi$
steradian area in the north Galactic cap (NGC).
For those times when the NGC is unavailable, about one third of the time,
we will repeatedly observe
three southern stripes, nominally centred at RA $\alpha = 5\dg$, and
with central declinations of $\delta = +15\dg$, $0\dg$ and $-10\dg$.
The nominal location of survey scans is shown in Figure~\ref{fig:stripes}.

In the remainder of this section I discuss the various components
of the survey in more detail.

{\bf 2.5-metre telescope.}
The main 2.5-metre telescope is of modified Richey-Chretien design
with a $3\dg$ field of view, and is optimised for both a wide-area
imaging survey and a multi-fibre spectroscopic survey of galaxies to
$r' \sim 18$.
One of the most unusual aspects of the telescope is it's enclosure.
Rather than sitting inside a dome, as is the case with conventional
optical telescopes, the enclosure is a rectangular frame structure
mounted on wheels, which is rolled away from the telescope in order
to take observations.
By completely removing the enclosure from the telescope, we can avoid
the substantial degradation to image quality due to dome seeing.
The telescope is situated on a pier overlooking a steep dropoff
so that the prevailing wind will flow smoothly over the telescope
in a laminar flow, which will also help to ensure good image quality.
A wind baffle closely surrounds the telescope, and is independently
mounted and driven.
This baffle serves to protect the telscope from stray light as
well as from wind buffeting.

\begin{figure}[htbp]
\parbox{10cm}{
\epsfxsize=10cm
\epsfbox{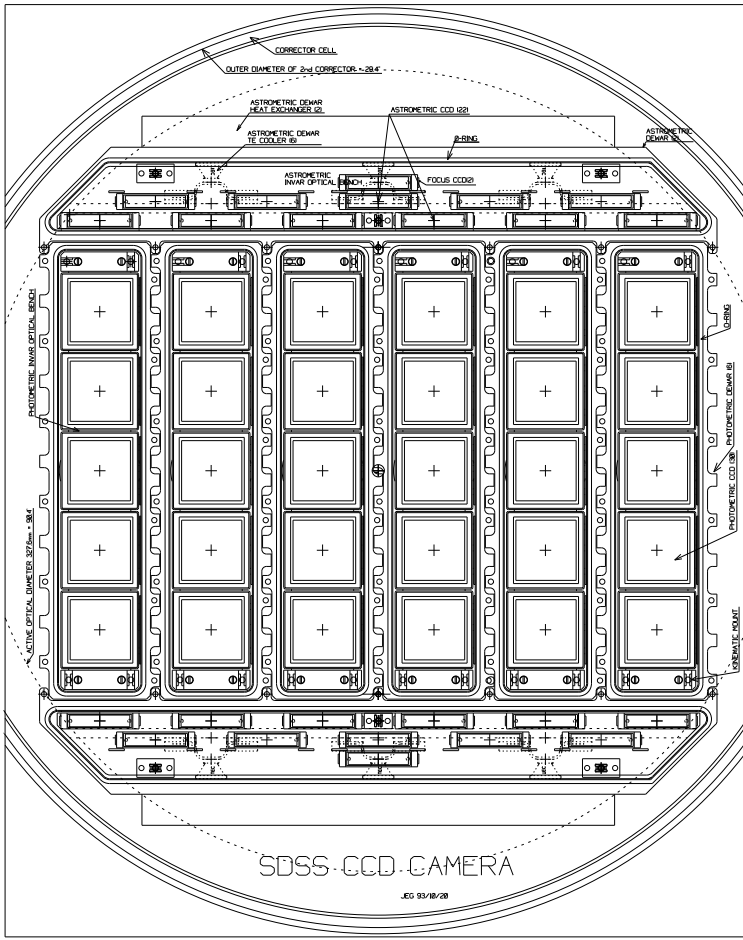}
}
\parbox{6cm}{
\caption[]{Focal plane layout of the SDSS CCD imaging camera, 
	showing the 30 photometric and 24
	astrometric/focus CCDs.}
\label{fig:imcam}
}
\end{figure}

{\bf Imaging Camera.}
In order to image a large area of sky in a short time, we are building
an imaging camera (Fig.~\ref{fig:imcam})
that contains $30 \times 2048^2$ CCDs, arranged in six columns.
Each column occupies its own dewar and contains one chip in each of
the five filters.  Pixel size is $0.4''$.
The camera operates in drift-scan mode: a star or galaxy image drifts down
the column through the five filters, spending about 55 seconds in each.
This mode of observing has two significant advantages over conventional
tracking mode.
1) It makes extremely efficient use of observing time, since there is no
overhead between exposures: on a good night we can open the shutter,
drift-scan for eight hours and then close the shutter.
2) Since each image traverses a whole column of pixels on each CCD,
flat-fielding becomes a one-dimensional problem, and so can be
done to lower surface-brightness limits than with tracking mode images.
This, along with the high quantum efficiency of modern CCDs, will enable us
to detect galaxies of much lower surface brightness than can wide-field
photographic surveys.
There is a gap between each column of CCDs, but this gap is slightly smaller
than the width of the light-sensitive area of the CCDs, and so having
observed six narrow strips of sky one night, we can observe an interleaving
set of strips a later night, and thus build up a large contiguous area of sky.
The northern survey comprises 45 pairs of interleaving great circle scans, 
and so
imaging observations for the north will require the equivalent of 90
full photometric nights.
The camera also includes 24 smaller CCDs arranged above and below the 
photometric columns.  These extra CCDs, equipped with neutral density filters,
are used for astrometric calibration,
as most astrometric standards will saturate on the photometric CCDs.
Thus the photometric data can be tied to the fundamental astrometric
reference frames defined by bright stars.

{\bf Spectrographs.}
The 2.5-metre telescope will also be equipped with a pair of fibre-fed,
dual-beam spectrographs, each with two cameras, two gratings and two $2048^2$
CCD detectors.
The blue channel will cover the wavelength range 3900--6100 \AA\ and
the red channel 5900--9100 \AA\ and both will have a spectral resolving
power $\lambda/\Delta \lambda \approx 1800$.
The fibres are $3''$ in diameter and the two spectrographs each hold 320 fibres.
Rather than employing robotic fibre positioners to place the fibres
in the focal plane, we will instead drill aluminium plates for each
spectroscopic field and plug the fibres by hand.
We plan on spectroscopic exposure times of 45 minutes and allow 15 minutes
overhead per fibre plate.
On a clear winter's night we can thus obtain 8 plates $\times$ 640 fibres
$= 5120$ spectra.
In order to allow such rapid turnaround time between exposures we plan
to purchase 8 sets of fibre harnesses, so that each plate can be plugged 
with fibres during the day.
It will not be necessary to plug each fibre in any particular hole,
as a fibre mapping system has been built which will automatically map
fibre number onto position in the focal plane after the plate has been plugged.
This should considerably ease the job of the fibre pluggers, and we expect
that it will take well under one hour to plug each plate.

{\bf Monitor telescope.}
In order to check that observing conditions are photometric,
and to tie imaging observations to a set of primary photometric standards,
we are also employing a monitor telescope.
While the 2.5-metre telescope is drift-scanning the sky,
the 0.6-metre monitor telescope, situated close by,
will interleave observations of standard stars with calibration patches
in the area of sky being scanned.
Operation of this telescope will be completely automated, and each hour will
observe three calibration patches plus standard stars in all five colours.

{\bf $10\mu$ all-sky camera.}
As an additional check on observing conditions, a $10\mu$ infrared
camera will survey the entire sky every 10 minutes or so.
Light cirrus, which is very hard to see on a dark night, is bright
at $10\mu$, and so this camera will provide rapid warning of
increasing cloud cover, thus enabling us to switch to spectroscopic
observing rather than taking non-photometric imaging data.

\begin{figure}[htbp]
\begin{center}
\leavevmode
\epsfxsize=0.6\textwidth
\epsfbox{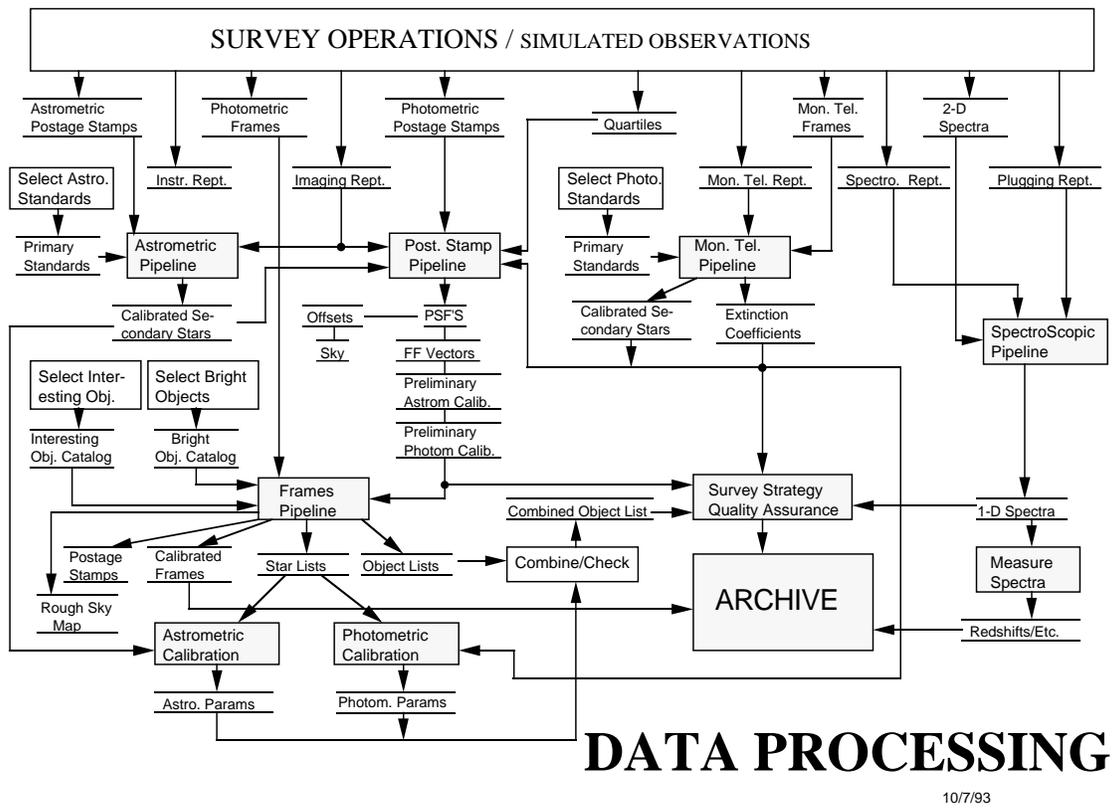}
\end{center}
\caption[]{Top-level data processing diagram.}
\label{fig:dataflow}
\end{figure}

{\bf Data-reduction pipelines.}
The last, but by no means least, component of the survey is a suite
of automated data-reduction pipelines (Fig.~\ref{fig:dataflow}), 
which will read DLT tapes mailed
to Fermilab from the mountain and yield reduced and calibrated data
with the minimum of human intervention.
Such software is very necessary when one considers that the imaging camera
will produce data at the rate of around 31 Gbytes per hour!
A ``production system'' has been specced and purchased that can keep up
with such a data rate (bearing in mind that imaging will take place
only under the best conditions, on average around two full nights per month),
and consists of two Digital Alphaserver 8200 5/300s, each with
1 GByte of memory.

Pipelines exist to reduce each source of data from the mountain
(photometric frames and ``postage stamps'', astrometric frames,
monitor telescope frames and 2-D spectra) as well as to perform
tasks such as spectroscopic target selection and ``adaptive tiling''
to work out the optimal placing of spectroscopic field centres to maximize
the number of spectra obtained.
The pipelines are integrated into a purpose-written environment
known as SHIVA (Survey Human Interface and VisualizAtion environment,
also the Hindu god of destruction)
and the reduced data will be written into an object-oriented database.

\section{Data Products}

The raw imaging data in five colours for the $\pi$ steradians of the 
northern sky will occupy about 14 Tbytes, but it is expected that very
few projects will need to access the raw data, which will probably
be stored only on magnetic tape.
Since most of the sky is blank to $r' \sim 23$, all detected images
can be stored, using suitable compression, in around 200 Gbytes,
and it is expected that these ``atlas images'' can be kept on spinning disc.
The photometric reduction pipeline will meaure a set of parameters
for each image, and it is estimated that the parameter lists for all objects
will occupy $\sim 100$ Gbyte.
The parameter lists for the spectroscopic sample will proabbly fit
into 1--2 Gb, and the spectra themselves will occupy $\sim 20$ Gb.
Work is progressing well on an astronomer-friendly interface to the
database, which will answer such queries as ``Return all galaxies
with $(g' - r') < 0.5$ and within 30 arcminutes of this quasar'', etc.

\subsection{Spectroscopic Samples}

The spectroscopic sample is divided into several classes.
In a survey of this magnitude, it is important that the selection 
criteria for each class remain fixed throughout the duration of the survey.
Therefore, we will spend a considerable time (maybe one year), obtaining
test data with the survey instruments and refining the spectroscopic selection
criteria in light of our test data.
Then, once the survey proper has commenced, these criteria will be
``frozen in'' for the duration of the survey.
The numbers discussed below are therefore only preliminary, and we expect
them to change slightly during the test year.

The {\bf main galaxy sample} will consist of $\sim 900,000$ galaxies selected
by Petrosian magnitude in the $r'$ band, $r' \simlt 18$.
Simulations have shown that the Petrosian magnitude,
which is based on an aperture defined by the ratio of light within an annulus 
to total light inside that radius, provides probably the least biased
and most stable estimate of total magnitude.
There will also be a surface-brightness limit, so that we do not
waste fibres on galaxies of too low surface brightness to give a reasonable
spectrum.
This galaxy sample will have a median redshift $\langle z \rangle \approx 0.1$.

We plan to observe an additional $\sim 100,000$
{\bf luminous red galaxies} to $r' \simlt 19.5$.
Given photometry in the five survey bands, redshifts can be estimated
for the reddest galaxies to $\Delta z \approx 0.02$ or better \cite{c95},
and so one can also predict their luminosity quite accurately.
Selecting luminous red galaxies, many of which will be cD galaxies in cluster 
cores, provides a valuable supplement to the main
galaxy sample since 1) they will have distinctive spectral features,
allowing a redshift to be measured up to 1.5 mag fainter than the main sample,
and 2) they will form an approximately volume-limited sample with a
median redshift $\langle z \rangle \approx 0.5$.
They will thus provide an extremely powerful sample for studying
clustering on the largest scales and the evolution of galaxies.

{\bf Quasar} candidates will be selected by making cuts
in multi-colour space and from the FIRST radio catalogue \cite{bwh95},
with the aim of observing $\sim 100,000$ quasars.
This sample will be orders of magnitude larger than any existing quasar
catalogues, and will be invaluable for quasar luminosity function, evolution
and clustering studies
as well as providing sources for followup absorption-line observations.

In addition to the above three classes of spectroscopic sources, which are
designed to provide {\em statistically complete} samples, we will also 
obtain spectra
for many thousands of {\bf stars} and for various {\bf serendipitous}
objects.
The latter class will include objects of unusual colour or morphology
which do not fit into the earlier classes, plus unusual objects found
by other surveys and in other wavebands.

\section{Current Status}

In this section I discuss the status (as of April 1996) 
of the various systems within the survey.

The {\bf monitor telescope} has been operational now for several months,
and is routinely operated remotely from Chicago.
It is equipped with a set of SDSS filters, and is being used to observe 
candidate primary photometric standard stars, as well as known quasars
to see where they lie in the SDSS colour system \cite{r96}.

\begin{figure}[htbp]
\parbox{10cm}{
\epsfxsize=10cm
\epsfbox{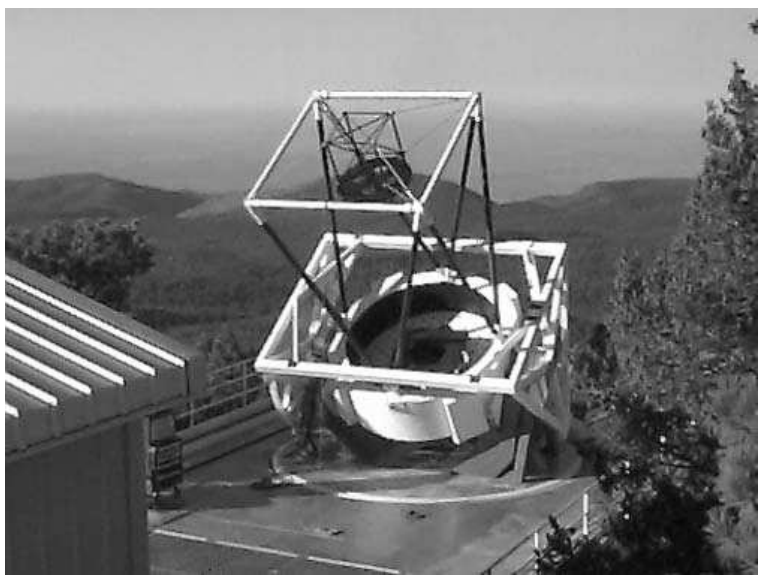}
}
\hspace{0.5cm}
\parbox{6cm}{
\caption[]{Photograph of the 2.5-metre telescope structure, taken shortly
	after installation, on 10 October 1995.
	Part of the telescope enclosure, in its rolled back position, appears
	in the bottom-left of this picture.
	Note that neither the mirrors nor the wind baffle are installed yet.}
\label{fig:tel}
}
\end{figure}

The {\bf 2.5-metre telescope} structure was installed on the mountaintop
in October 1995 (see Fig.~\ref{fig:tel}).
Work is currently underway on the control systems for the telescope.
Telescope {\bf optics} are all due to be ready by June 1996.
These include the primary and secondary mirrors and various corrector elements.

We posess all of the CCDs for the {\bf imaging camera}, which is under 
construction at Princeton.
Delivery to the mountain is expected by September 1996.
Construction of the {\bf spectrographs} is well underway, with the
optics installed for one of the spectrograph cameras.

Each of the data reduction-pipelines is now basically working, with 
ongoing work on minor bug-fixes, speed-ups and integration of the entire
data processing system.
The photometric reduction pipeline is being tested using both simulated data
and with data taken using the Fermilab drift scan camera on the ARC 3.5-metre
telescope at the same site.
Similar tests are being carried out on the spectroscopic reduction pipeline,
and our ability to efficiently place fibres on a clustered distribution of 
galaxies is being tested using the APM galaxy catalogue \cite{msel90}.

The currently-projected survey schedule is as follows:

\vspace{0.5cm}
\begin{tabular}{ll}
September 1996 & Optics to be installed on 2.5-metre telescope.\\
Autumn 1996 & Imager and spectrograph commissioning.\\
Winter 1996 & Astronomical first light.\\
Early 1997 & Test period begins.\\
1998--2003 & Survey proper carried out.\\
2002 & First two years of survey data become public.\\
2005 & Complete survey data become public.\\
\end{tabular}
\vspace{0.5cm}

The intent of this project is to make the survey data available to the
astronomical community in a timely fashion. We currently plan to
distribute the data from the first two years of the survey no later
than two years after it is taken, and the full survey no later than
two years after it is finished.  The first partial release may or may
not be in its final form, depending on our ability to calibrate it
fully at the time of the release. The same remarks apply to the
release of the full data set, but we expect the calibration effort to
be finished before that release.

\section{Prospects for constraining dark matter}

Since one of the topics of this meeting is dark matter, I will highlight
two of the areas in which the SDSS will provide valuable data for
constraining dark matter.

\subsection{Measurement of the Fluctuation Spectrum}

\begin{figure}[htbp]
\parbox{10cm}{
\epsfxsize=10cm
\epsfbox{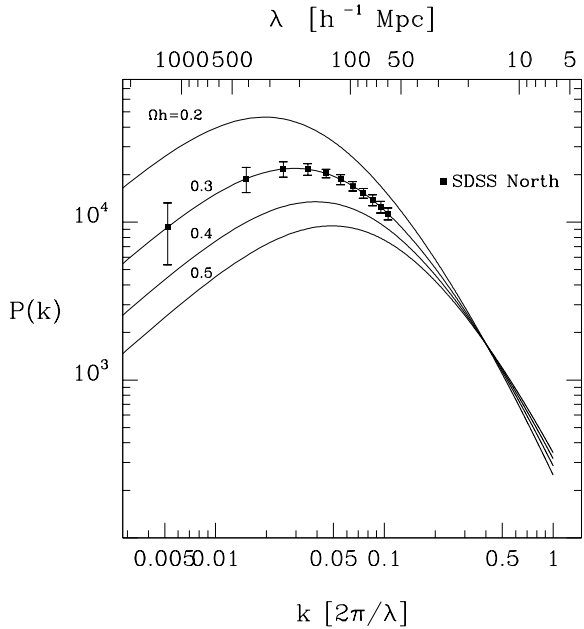}
\vspace{-1cm}
}
\parbox{6cm}{
\caption[]{Expected $1\sigma$ uncertainty in the galaxy power spectrum
	measured from a volume-limited sample from the SDSS northern survey,
	along with predictions of $P(k)$ from four variants of the
	low-density CDM model.
	Note that the models have been arbitrarily normalised to agree
	on small scales ($k = 0.4$); in practice the COBE observations
	of CMB fluctuations fix the amplitude of $P(k)$ on very large scales.}
\label{fig:P_k}
}
\end{figure}

The huge volume of the SDSS redshift survey will enable
estimates of the galaxy power spectrum to $\sim 1000 \hMpc$ scales.
Figure~\ref{fig:P_k} shows the power spectrum $P(k)$ we would expect
to measure from a volume-limited (to $M^*$) sample of galaxies from
the SDSS northern redshift survey, assuming Gaussian fluctuations
and a $\Omega h = 0.3$ CDM model.
The error bars include cosmic variance and shot noise, but not systematic
errors, due, for example, to galactic obscuration.
Provided such errors can be corrected for, (and star colours in the Sloan
survey will provide our best {\em a posteriori}
estimate of galactic obscuration),
then the figure shows that
we can easily distinguish between $\Omega h = 0.2$ and $\Omega h = 0.3$ models,
just using the northern main galaxy sample.
Adding the southern stripe data, and the luminous red galaxy sample, will
further decrease measurement errors on the largest scales, and so we also
expect to be able to easily distinguish between low-density CDM and MDM models,
and models with differing indices $n$ for the shape of the primordial
fluctuation spectrum.

\subsection{Cosmological Density Parameter}

By measuring the distortions introduced by streaming motions
into redshift-space measures of galaxy clustering, one can constrain the
parameter $\beta = \Omega^{0.6}/b$, where $\Omega$ is the cosmological 
density paramter and $b$ is the bias factor relating fluctuations in
galaxy number density to fluctuations in the underlying mass distribution.
While existing redshift surveys, eg. IRAS \cite{cfw95} and Stromlo-APM
\cite{lemp96}, are hinting
that $\beta < 1$ (ie. that galaxies are significantly biased tracers
of mass or that $\Omega < 1$), their volumes are too small to
measure galaxy clustering in the fully linear regime reliably enough
to measure $\beta$ to much better than 50\% or so.
With the SDSS redshift survey, we expect to be able to constrain
$\beta$ to 10\% or better.

There are several ways we might hope to determine the galaxy bias factor $b$.
By measuring galaxy clustering on $\sim 1000 \hMpc$ scales as shown in
Figure~\ref{fig:P_k}, we can compare
with the COBE microwave background fluctuations directly, and so
constrain large-scale galaxy bias in a model-independent way.
Analysis of higher-order clustering statistics \cite{gf94},
and of non-linear dynamical effects \cite{cfw95} will also set constraints
on galaxy bias.
Knowing $\beta$ and $b$. we will be in a good position to reliably
measure the cosmological density parameter
$\Omega$ independent of models for the shape of the fluctuation spectrum.

\section{Conclusions}

It is probably no exaggeration to claim that the Sloan Digital Sky Survey
will revolutionize the field of large scale structure.
Certainly we can expect to rule out large numbers of presently viable 
cosmological models, as illustrated in Figure~\ref{fig:P_k}.
As well as measuring redshifts for a carefully controlled sample of
$10^6$ galaxies and $10^5$ quasars, the survey will also provide high quality
imaging data for about 100 times as many extragalactic objects,
from which one can obtain colour and morphological information.
In addition to measuring the basic cosmological parameters $\Omega$ and $h$
discussed in the preceding section, the SDSS will also allow us to measure the 
properties of galaxies as a function
of their colour, morphology and environment, providing valuable clues
to the process of galaxy formation.

\begin{figure}[htbp]
\parbox{10cm}{
\epsfxsize=10cm
\epsfbox{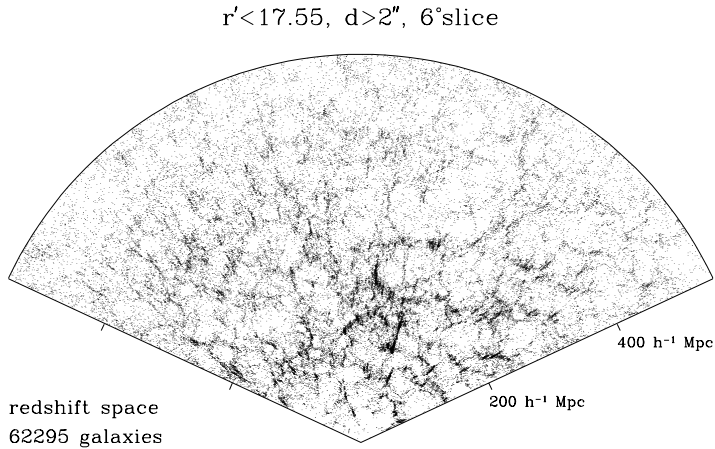}
}
\parbox{6cm}{
\caption[]{Redshift-space distribution of galaxies in a $6\dg$ slice from
	a large, low-density CDM $N$-body simulation generated by Changbom 
	Park.}
\label{fig:slice}
}
\end{figure}

Finally, I cannot resist the temptation to give a visual impression
of what we might expect to see with the SDSS redshift survey.
Figure~\ref{fig:slice} shows the distribution of 62,295 galaxies in
a $6\dg$ slice from a simulation carried out by Changbom Park,
assuming a low-density CDM model.
This slice represents just {\em one sixteenth} of the million galaxy
redshifts we will be measuring with the Sloan survey.
I leave it to the readers imagination to dream up all the projects
they would love to carry out given such a data-set.

The work described here has been carried out by many people throughout
the SDSS collaboration, and I thank all my colleagues warmly.
I am particularly grateful to Chris Stoughton and Michael Vogeley for providing
Figures~\ref{fig:stripes} and Figure~\ref{fig:P_k} respectively,
and to Philippe Canal for translating the Abstract into French.
My attendance at the meeting was supported by a generous grant from
the EEC.

\vspace{1cm}
\centerline{\bf LE SLOAN DIGITAL SKY SURVEY: 
L'\'ETAT ET CES PERSPECTIVES}
\vspace{0.5cm}

Le Sloan Digital Sky Survey (SDSS) \`a pour but de cartographi\'e
$\pi$ steradians de l'univers local.  Une matrice de dispositif \`a
transfert de charges (CCD) scannant en mode balayage produira une image
digitalis\'ee du ciel avec cinq diff\'erents filtres et avec une pr\'ecision
allant jusqu'a \`a peu pr\`es magnitude 23.  Une \'etude spectroscopique sera
faite sur une s\'election de $10^6$ galaxies et $10^5$ quasars.
Dans cet article, apr\`es avoir d\'ecris l'\'etat d'advancement du projet
qui doit commencer \`a faire des observations des le d\'ebut de l'ann\'ee
1997, je pr\'esente ces perspectives pour l'\'etablissement de mod\`eles de
la mati\`ere noire dans l'univers.


\begin{thebibliography}{99}{\baselineskip 0.4cm
\bibitem{bwh95}Becker, R.H., White, R.L. and Helfand, D.J., 1995, ApJ, 450, 559
\bibitem{cfw95}Cole, S., Fisher, K.B. and Weinberg, D.H., 1995, MNRAS, 275, 515
\bibitem{chm89}Collins, C.A., Heydon-Dumbleton, N.H. and  MacGillivray, H.T.,
1989, MNRAS, 236, 7P
\bibitem{c95}Connolly, A.J., et al., 1995, AJ, 110, 2655
\bibitem{gf94}Gazta\~{n}aga, E. and Frieman, J.A., 1994, ApJ, 437, L13
\bibitem{lemp96}Loveday, J., Efstathiou, G., Maddox, S.J. and Peterson, B.A.,
1996, ApJ, in press
\bibitem{msel90}Maddox, S.J., Sutherland, W.J. Efstathiou, G. and Loveday, J.,
1990, MNRAS, 243, 692
\bibitem{r96}Richards, G.T., et al., PASP, submitted
}
\end{thebibliography}
\end{document}